\documentclass[conference]{IEEEtran}
\usepackage{ifpdf}
\usepackage{cite}
\ifCLASSINFOpdf
   \usepackage[pdftex]{graphicx}
   \graphicspath{{../pdf/}{../jpeg/}}
   \DeclareGraphicsExtensions{.pdf,.jpeg,.png}
\else
   \usepackage[dvips]{graphicx}
   \graphicspath{{../eps/}}
   \DeclareGraphicsExtensions{.eps}
\fi
\usepackage{epsf,psfrag,epsfig}

\usepackage[cmex10]{amsmath}
\usepackage{mathrsfs, amsthm,amsfonts, latexsym, amssymb,color}
\usepackage{algorithmic,algorithm}
\usepackage[english]{babel}
\usepackage[mathscr]{eucal}
\usepackage{graphicx}
\usepackage[breaklinks]{hyperref}
\usepackage[hyphenbreaks]{breakurl}

{\theoremstyle{definition}
\newtheorem{defn}{Definition}}
{\theoremstyle{plain}

\newtheorem{thm}{Theorem}}

\IEEEoverridecommandlockouts
\renewcommand{\thefootnote}{\arabic{footnote}}

\def\nn{\nonumber}
\def\beq{\begin{equation}}
\def\eeq{\end{equation}}
\def\bea{\begin{eqnarray}}
\def\eea{\end{eqnarray}}
\def\ba{\begin{array}}
\def\ea{\end{array}}
\def\defeq{{\stackrel{\Delta}{=}}}

\def\bitem{\begin{itemize}}
\def\eitem{\end{itemize}}
\def\benum{\begin{enumerate}}
\def\eenum{\end{enumerate}}
\def\beq{\begin{equation}}
\def\eeq{\end{equation}}
\def\bea{\begin{eqnarray}}
\def\eea{\end{eqnarray}}

\def\etal{{\it et al. \/}}

\definecolor{bgrd}{rgb}{1,1,1}
\definecolor{gray}{rgb}{0.5,0.5,0.5}
\definecolor{dkr}{rgb}{0.7,0.1,0.2}
\definecolor{dkb}{rgb}{0.1,0.1,0.8}

\makeatletter
\newdimen{\captionwidth}
\long\def\@makecaption#1#2{%
\captionwidth .9\hsize
\vskip 10pt%
\setbox\@tempboxa\hbox{#1: #2}%
  \ifdim \wd\@tempboxa >\captionwidth%
    \setbox\@tempboxa\hbox{#1:\hspace*{.5em}}%
    \hfil\parbox{\captionwidth}{\raggedright\hangindent \wd\@tempboxa%
    \hangafter=1\unhbox\@tempboxa#2}\hfill%
  \else\centerline{\box\@tempboxa}%
  \fi
}
\makeatother

\def\edoc{\end{document}}

\def\diag{\mbox{diag}}

\newcommand{\Cmsc}{\mathscr{C}}

\newcommand{\Pmsc}{\mathscr{P}}

\def\Nc{{\cal N}}

\begin{document}

\title{Probabilistic Forecasting and Simulation of  Electricity Markets via Online Dictionary Learning}

\author{
\IEEEauthorblockN{\Large Weisi Deng, Yuting Ji,  and Lang Tong}
}

\maketitle{
\let\thefootnote\relax\footnotetext{
This work is supported in part by  a Grant from DoE Consortium for Electric Reliability Technology Solutions (CERTS), the National Science Foundation, and the Chinese Government Graduate Student Overseas Study Program.\\ $\dagger$ The authors are alphabetically ordered.  Weisi Deng is with the School of Electrical and Electronic Engineering, Huazhong University of Science and Technology, Wuhan, China. Yuting Ji and Lang Tong are with the School of Electrical and Computer Engineering, Cornell University, Ithaca, NY, USA. Email: \{wd229,yj246,lt35\}@cornell.edu}
}	

\begin{abstract}
The problem of probabilistic forecasting and online simulation of real-time electricity market with stochastic generation and demand is considered.   By exploiting the parametric structure of the direct current optimal power flow, a new technique  based on online dictionary learning  (ODL)  is proposed. The ODL approach incorporates real-time measurements and historical traces to produce forecasts of joint and marginal probability distributions of future locational marginal prices, power flows, and dispatch levels, conditional on the system state at the time of forecasting.  Compared with standard Monte Carlo simulation techniques, the ODL approach offers several orders of magnitude improvement in computation time, making it feasible for online forecasting of market operations.  Numerical simulations on large and moderate size power systems illustrate its performance and complexity features and its potential as a  tool for system operators.

\end{abstract}

\begin{IEEEkeywords}
Electricity market, locational marginal price (LMP), probabilistic price forecasting, power flow distributions, dictionary learning, machine learning in power systems.
\end{IEEEkeywords}

\IEEEpeerreviewmaketitle

\section{Introduction}
We consider the problem of online forecasting and simulation of real-time  wholesale electricity market.  By online forecasting and simulation we mean in particular the use of real-time SCADA and PMU measurements to produce conditional probability distributions of future nodal prices, power flows, power dispatch levels, and discrete events such as congestions and occurrences of contingencies.

The forecasting and simulation problem considered in this paper is motivated by the increasing presence of stochastic elements in power system as a result of integrating  intermittent renewables at both  wholesale and retail levels.   The surge of solar power integration in recent years, for example,  has fundamentally changed the overall netload characteristics.  In some areas,  the traditional  load profile is being transformed to the so-called ``duck curve'' profile where  a steep down-ramp during the hours when a large amount of solar power is injected into the network is followed by a steeper up-ramp when the solar power drops sharply in the late afternoon and early evening hours.

While the duck curve phenomenon represents an {\em average} net load behavior, it is the highly stochastic and spatial-temporal dependent ramp events  that  present difficult operational challenges to system operators.  For this reason, there is a need for a more detailed and informative characterization of the overall uncertainty of power system operation, one that  reveals interdependencies of power flows, congestions, and locational marginal prices (LMPs).

Currently,  some system operators are providing real-time price forecasts. The Electric Reliability Council of Texas (ERCOT)\cite{ercot} offers one-hour ahead real-time LMP forecasts, updated every 5 minutes. Such forecasts signal potential shortage/oversupply caused by anticipated fall/rise of  renewable supplies or the likelihood of network congestions.   The Alberta Electric System Operator (AESO) \cite{aeso} provides two short-term price forecasts with prediction horizons of 2 hours and 6 hours, respectively.

  Most LMP forecasts, especially those provided by system operators,  are {\em point forecasts} that predict directly future LMP values.  They are typically generated  by substituting the  expected trajectory of random load and intermittent generation in place of their actual realizations.  Such  {\em certainty equivalent} approaches amount to equating the expectation of a function of random variables and a function of the expectation of random variables; they can lead to gross mischaracterization of the behavior of the system operation.  More significant, perhaps, is that  point forecasts are of limited value if forecasts are to be integrated into system and market operations.   To a system operator as well as market participants, the most informative type of forecasting---one that is the focus of this paper---is the  {\em probabilistic forecasting} that produces  probability distribution of future system variables conditional on the current system state.

\subsection{Related Work}
There is a substantial literature on point forecasting techniques from the perspectives of external market participants.  See \cite{Weron:14} for a recent review.  These techniques do not incorporate system operating conditions that are only available to system operators.  Here we highlight some results on probabilistic forecasting from the system operator's perspectives.

Probabilistic forecasting has not been widely used in power system operations because of the difficulty associated with obtaining conditional probability distributions of future system variables.  Other than some simple cases, probabilistic forecasting in a large complex system can only be obtained by Monte Carlo techniques where conditional distributions are estimated using sample paths  generated either according to the underlying system model or directly from measurements and historical data.  In this context,  the problem  obtaining probabilistic forecasting is essentially the same as  online Monte Carlo simulations.   To this end, there is a premium on the cost computation and the rate of convergence of statistics.

There are several prior approaches to probabilistic forecasting for system operators.   In particular, Min \etal proposed in \cite{Min08}  direct implementation  of  Monte Carlo simulations to obtain short-term forecasting of transmission congestion.  For  $M$ Monte Carlo runs over a $T$-period  forecasting horizon, the computation cost is dominated by the computation of  $M\times T$  direct current optimal power flow (DCOPF) solutions that are used to generate the necessary statistics.  For a large scale system with a significant amount of random generations and loads, such computation costs may be too high for such a technique to be used for online forecasting.

A similar approach based on a nonhomogeneous Markov chain modeling of real-time LMP is proposed in \cite{Ji13Asilomar}.  The Markov chain technique  exploits the discrete nature of LMP distributions and obtains LMP forecasts by the product of transition matrices of LMP states.  Estimating the transition probabilities, however, requires roughly the same number of Monte Carlo simulations, thus requiring roughly the same number of DCOPF computations.

An existing work closest to the present paper is \cite{JiThomasTong15HICSS_LMP,JiThomasTong15ArXiv_LMP} where the authors proposed a probabilistic forecasting method based on a multiparametric formulation of DCOPF that has random generations and demands as parameters.   From the parametric linear/quadrtic programming theory, the (conditional) probability distributions of LMP and power flows, given the current system state, reduce to the conditional probabilities that  realizations of the random demand and generation fall into one of the  critical regions in the parameter space.

The main difficulty of the approach in \cite{JiThomasTong15HICSS_LMP,JiThomasTong15ArXiv_LMP}  is the high cost of computing the set of critical regions that partition the parameter space.  Although such computations can be made off-line,  the number of critical regions grows exponentially with the number of constraints, which makes  even the off-line computations prohibitive for large systems.   The approach presented in this paper builds upon the ideas in \cite{JiThomasTong15HICSS_LMP} and develops a computationally efficient  and adaptive forecasting technique.

Among techniques applicable to forecasting problems by system operators,  several are based on approximations of future LMP distributions. In \cite{Bo09TPS}, a probabilistic LMP forecasting approach was proposed based on attaching a Gaussian distribution to a point estimate.  While the technique can be used to generalize point forecasting methods to probabilistic ones, it has the limitation in incorporating network effects.   The authors of \cite{mokhtari2009new} and \cite{davari2008determination}  approximate the probabilistic distribution of LMP using higher order moments and cumulants.  These methods are based on representing the probability distribution as an infinite series involving moments or cumulants. In practice, computing or estimating higher order moments and cumulants are very difficult; lower order approximations are necessary.

\subsection{Summary of Contribution}
In this paper, we present a new methodology for the probabilistic forecasting and online simulation of real-time electricity market.  The main idea  is  {\em online dictionary learning} (ODL) that sequentially captures the parametric structure of DCOPF solutions.  The main features of the proposed methodology are the significant reduction of computation costs and its ability of adapting to changing operating conditions.  For large systems, the ODL approach offers several orders of magnitude improvement in computation cost.

The ODL approach is a Monte Carlo simulation method  with two key innovations.  First, the proposed approach is based on a multiparametric DCOPF formulation for real-time market operations. By exploiting explicitly the solution structure of DCOPF, we reduce the problem of collecting statistics on the space of {\em continuous probability distributions} of random parameters (generation and demand) to the space of {\em finite discrete probability distributions} on a set of critical regions.  Note that each critical region is associated with a unique affine function that maps the parameter to the solution of DCOPF and the associated Lagrange multipliers.

Second, we propose an online dictionary learning approach that sequentially builds a dictionary of solutions from past samples using a dynamic critical region generation  process.  In particular, each entry of the dictionary corresponds to an {\em observed} critical region within which a sample of random generation/demand has fallen.   A new entry of the dictionary is produced only when the realization of the renewable generation and demand does not fall into one of the existing critical regions  in the dictionary.  This allows us to avoid  costly DCOPF  computations and recall directly the solution from the dictionary.  Because renewable generation and load processes are physical processes, they tend to be bounded and  concentrated around the mean trajectory.  As a result, despite that there are potentially exponentially large number of potential entries in the dictionary, only a very small fraction of the dictionary entries are observed in the simulation process.

\section{Parametric Models of Real-Time Operation}\label{sec:dcopf}
Most wholesale	electricity market \cite{PJM, NYISO, ISONE} consists of day-ahead  and real-time markets. The day-ahead market enables market participants commit to buy or sell wholesale electricity one day before operation, and the real-time market balances the differences between day-ahead commitments and the actual real-time demand and production. In this paper, we focus on real-time operation models.  In particular, we consider two real-time markets: one is the  energy-only market; the other is the co-optimized energy-reserve market.  Our approach also applies to several other real-time markets such as the capacity and ancillary service markets.

Our presentation here highlights a {\em parametric} formulation that treats random elements in the system such as renewable generation, demands, etc., as parameters that vary from  time to time and realization to realization.

\subsection{Energy Only Market}
In the energy-only market, the operator sets generation adjustments by solving a DCOPF problem in which the one-step ahead real-time demand is balanced subject to system constraints \cite{Ott03TPS_ED}. For simplicity, we assume that each bus has a generator and a load. The DCOPF problem at time $t$ is defined by the following optimization:
\begin{equation}\label{eqn:dcopf}
\begin{array}{l l l}
\underset{g}{\text{minimize}}& c(g)& \\
\text{subject to}&\\
& \mathbf{1}^\intercal (g-d_{t})=0 &(\lambda_t) \\
& -F^+ \leq S(g-d_{t}) \leq F^+&(\mu_t^+, \mu_t^-) \\
&g^-\leq g \leq g^+ & \\
& \hat{g}_{t-1}-\Delta^{-} \leq g \leq \hat{g}_{t-1}+\Delta^{+}	
\end{array}
\end{equation}
where\\
\begin{tabular}{p{.9cm} p{7.36cm}}
$c(\cdot)$&real-time generation cost function;\\
$g$&vector of ex-ante dispatch at time $t$;\\
$\hat{g}_{t-1}$&vector of generation estimate at time $t-1$;\\
$d_{t}$& vector of one-step net load\footnotemark at time $t$;\\
$F^+/F^-$&vector of max/min transmission capacities;\\
$g^+/g^-$&vector of  max/min generator capacities;\\
$\Delta^+/\Delta^-$&vector of  upward/downward ramp limits;\\
$S$& shift factor matrix;\\
$\lambda_t$& shadow price for the energy balance constraint at time $t$;\\
$\mu_t^+/\mu_t^-$&shadow prices for max/min transmission constraints at time $t$.
\end{tabular}
In this model,  the generation costs can be linear, piece-wise affine, or strictly convex quadratic.   The real-time LMP $\pi_{t}$ at time $t$ are calculated from the (dual) solutions of  (\ref{eqn:dcopf}) as the sum of the energy and congestion prices
\begin{equation}\label{lmp1}
\pi_{t}=\lambda_t {\mathbf{1}} -S^\intercal\mu_t^+ +S^\intercal\mu_t^-.
\end{equation}

Given the predicted load $d_t$ and estimated  generation (from SCADA or PMU measurements) $\hat{g}_{t-1}$, the above optimization can be viewed as a {\em parametric  DCOPF} with  parameter $\theta=(d_t, \hat{g}_{t-1})$.
 This viewpoint plays a critical role in our approach.

\footnotetext{In this model, we use the concept of ``net load'' $d_t$. Since renewable generation can be considered as a negative load, we define the net load as the total electrical load plus interchange minus the renewable generation. The interchange schedule refers to the total scheduled delivery and receipt of power and energy of the neighboring areas.}

\subsection{Joint Energy and Reserve Market}
In the joint energy and reserve market, dispatch and reserve  are jointly determined via a linear program that minimizes the overall cost subject to  operating constraints. In the co-optimized energy and reserve market, system-wide and locational reserve constraints are enforced by the market operator to procure enough reserves to cover the first and the second
contingency events. We adopt the co-optimization model in \cite{ZhengLitvinov06TPS_Cooptimization} as follows:

\begin{equation}\begin{array}{l l l}
\underset{g,r,s}{\text{minimize}} & \multicolumn{2}{l} {\sum_i \left(c^g_i g_i + {\sum_j} c^{r}_{i,j} r_{i,j}\right)
+{\sum_u} c^{p}_u s_u^{l} +{\sum_v} c^{p}_v s_v^{s}}\\
\text{subject to}&&\\
& \mathbf{1}^\intercal (g-d)=0 &\\
& -F^+ \leq S(g-d) \leq F^+&\\
& {\sum_i}{\sum_j} \delta^u_{i,j} r_{i,j}+(I_u^+-I_u)+s_u^{l} \geq Q^{l}_u &\forall u\\
&I_u={\sum_{i}} {\sum_{k \in I_u}}S_{ik} (g_i-{d_i}) &\\
& {\sum_i} {\sum_j} \delta^v_{i,j}r_{i,j}+s_v^{s} \geq Q^{s}_v &\forall v\\
&  g_i^-\leq {{g}_i}\leq g_i^+-{\sum_j}r_{i,j} &\forall i\\
& { \hat{g}_{t-1}}-\Delta^{-} \leq g \leq {\hat{g}_{t-1}}+\Delta^{+} &\\
& 0 \leq r \leq r^+ &\\
& s_u^l, s_v^s \geq 0 &\forall u, v\\
\end{array}\end{equation}
where\\
\begin{tabular}{p{.9cm} p{7.36cm}}
$i$ &index of buses;\\
$j$ &index of reserve types, 10-min spinning, 10-min non spinning, or 30-min operating reserve;\\
$u/v$ &index of locational/system-wide reserve constraints;\\
$k$ &index of transmission constraints;\\
$d_i$& net load at bus $i$;\\
$g_i$ & dispatch of online generator at bus $i$;\\
$\hat{g}_{t-1}$&vector of generation estimate at time $t-1$;\\
$r_{i,j}$&generation reserve of type $j$ at bus $i$;\\
$s^l/s^s$ &local/system reserve deficit of constraint;\\
$c^g_i$& cost for generation at bus $i$;\\
$c^{r}_{i,j}$& cost for reserve type $j$ at bus $i$;\\
$c^{p}_{u/v}$& penalty for reserve deficit of constraint $u/v$;\\
$I_u$ &interface flow for locational reserve constraint $u$;\\
$I_u^+$ &interface flow limit for locational reserve constraint $u$;\\
$F^+/F^-$&vector of max/min transmission capacities;\\
$Q^l/Q^s$&locational/system reserve requirement of constraint;\\
$S$& shift factor matrix;\\
$\delta_{i,j}^x$&binary value that is 1 when reserve $j$ at bus $i$ belongs to constraint $x$;\\
$g_i^{+/-}$& max/min generation capacity for generator at bus $i$;\\
$\Delta^{+/-}$& vector of upward/downward ramp limits;\\
$r^+$ & vector of ramp capacities.\\
\end{tabular}

Note again that the energy-reserve co-optimization model is also of the form of parametric DCOPF with parameter $\theta=(d_t, \hat{g}_{t-1})$ that is realized prior to the co-optimization.

\section{Multiparametric Program}\label{sec:multiparametric}
We have seen in previous section that a number of real-time market operations can be modeled in the form of parametric DCOPF. In this section, we summarize several key results on multiparametric linear/quadratic programming that we use to develop our approach.  See \cite{Gal72,BemporadEtal02Automatica_MQP,BorrelliBemporadMorari03JOTA_MLP, mpc_book} for multiparametric programming for more comprehensive expositions.

Consider a general right-hand side\footnote{By right-hand side we mean the parameter vector $\theta$ is on the right-hand side of the constraint inequalities.} multiparametric program (MPP) as follows:
\begin{equation}\label{eqn:mp}
\begin{array}{c}
\underset{x}{\text{minimize}}~z(x) \text{ subject to } Ax \leq b+ E\theta \quad (y)\\
\end{array}\end{equation}
where $x$ is the decision vector,  $\theta$ the parameter vector, $z(\cdot)$ the cost function, $y$ the Lagrangian multiplier vector, and $A$, $E$, $b$ are coefficient matrix/vector with compatible dimensions.

The multiparametric programming problem is to solve  (\ref{eqn:mp}) for all values of the parameter vector $\theta$: the optimal primal solution $x^*(\theta)$, the associated dual solution $y^*(\theta)$, and the the value of optimization $z^*(\theta)$.

In this paper, we only consider the linear and quadratic programs for which the multiparametric programming problems are referred to as multiparametric linear programs (MPLP) and multiparametric quadratic programs (MPQP) respectively. In addition, we assume that the MP is not (primal or dual) degenerate\footnote{For a given $\theta$, the MP (\ref{eqn:mp}) is said to be primal degenerate if there exists an optimal solution $x^*(\theta)$ such that the number of active constraints is greater than the dimension of $x$. By dual degeneracy we mean that the dual problem of MPP (\ref{eqn:mp}) is primal degenerate.} for all parameter values.  Under this assumption, the primal and dual solutions to (\ref{eqn:mp}) are unique for all $\theta$. Approaches for the degeneracy cases are presented in \cite{mpc_book}.

\subsection{Critical Region and Solution Structure}
The multiparametric programming analysis and the proposed simulation technique build upon the concept of {\em critical region.} Critical region partitions the parameter space into a finite number of regions.  Within each critical region, there is an affine relation between parameter value and optimization solution.

There are several definitions for critical region, we adopt the definition from \cite{mpc_book} under primal/dual  non-degeneracy assumption.
\begin{defn}\label{def:critical_region}
A \textit{critical region} $\Theta$ is defined as the set of all parameters such that for every pair of parameters $\theta, \theta' \in \Theta$, their respective solutions  $x^*(\theta)$ and $x^*(\theta')$ of (\ref{eqn:mp}) have the same active/inactive constraints.
\end{defn}

The definition  implies that each critical region is a polyhedron in the parameter space.  Given an MPP (\ref{eqn:mp}), the set of critical regions can be computed explicitly, although the cost of constructing of the complete set of critical regions may grow exponentially with the number of constraints.

In this paper, we avoid computing the complete set of critical regions.  Instead, we dynamically generate critical regions on demand.  To this end, we need a procedure to compute the critical region that contains a given parameter and the mapping of the parameter to  the primal and dual solutions of (\ref{eqn:mp}).  These results are summarized in the following theorem.

\begin{thm}\label{thm:cr}
  Consider (\ref{eqn:mp}) with  cost function $z(x)=c^\intercal x$ for MPLP and $z(x)=\frac{1}{2}x^\intercal H x$ for MPQP   where $H$ is  positive definite. Given parameter $\theta_0$ and the solution of the parametric program $x^*(\theta_0)$,   let $\tilde{A}, \tilde{E}$ and $\tilde{b}$ be, respectively, the submatrices of $A$, $E$ and subvector of $b$ corresponding to the active constraints.  Let $\bar{A}, \bar{E}$ and $\bar{b}$ be similarly defined for the inactive constraints.  Assume that (\ref{eqn:mp}) is neither  primal nor dual degenerate.
 \begin{enumerate}
 \item[(1)] For the MPLP, the critical region $\Cmsc_0$ that contains $\theta_0$ is given by, respectively,
  \bea
  \Cmsc_0&=&\big\{\theta\big| (\bar{A}\tilde{A}^{-1}\tilde{E}-\bar{E})\theta < \bar{b}-\bar{A}\tilde{A}^{-1}\tilde{b}\big\}.
  \eea
  And for any $\theta \in \Cmsc_0$, the primal and dual solutions are given by
  \[
  x^*(\theta)=\tilde{A}^{-1}(\tilde{b}+\tilde{E}\theta),~~y^*(\theta)=y^*(\theta_0).
  \]

  \item [(2)] For the MPQP, the critical region $\Cmsc_0$ that contains $\theta_0$ is given by
   \beq
   \Cmsc_0=\{\theta| \theta\in \Pmsc_p  \bigcap \Pmsc_d \}
   \eeq
   where $\Pmsc_p$  and  $\Pmsc_d$ are polyhedra defined by
\end{enumerate}
\[\begin{array}{l}\Pmsc_p=\{\theta |\bar{A}H^{-1}\tilde{A}^\intercal(\tilde{A}H^{-1}\tilde{A}^\intercal)^{-1}(\tilde{b}+\tilde{E}\theta)-\bar{b}-\bar{E}\theta < \mathbf{0} \}\\
\Pmsc_d=\{\theta|(\tilde{A}H^{-1}\tilde{A}^\intercal)^{-1}(\tilde{b}+\tilde{E}\theta)\le \mathbf{0}\}.\end{array}\]
 And for any $\theta \in \Cmsc_0$, the primal and dual solutions are given by
\bea
x^*(\theta) &=& H^{-1}\tilde{A}^\intercal(\tilde{A}H^{-1}\tilde{A}^\intercal)^{-1}(\tilde{b}+\tilde{E}\theta)\nn\\
y^*(\theta) &=&\left\{\begin{array}{ll}
0 & \mbox{inactive constraints}\\
 -(\tilde{A}H^{-1}\tilde{A}^\intercal)^{-1}(\tilde{b}+\tilde{E}\theta)\nn & \mbox{active constraints}
 \end{array}
 \right..
\eea

\end{thm}

\vspace{1em}
The proof of above theorem follows some of the derivations in \cite{mpc_book} (Chapter 7) and is consolidated in the Appendix.

For our application, a key implication of this theorem is that, once we know that a realized random parameter $\theta$ is in a known critical region, we no longer need to solve the original LP/QP; the solutions can be easily computed from the affine mappings.

\section{Forecasting  via  Online Dictionary Learning}\label{sec:algorithm}
We present in this section a new methodology of probabilistic forecasting and online simulation of real-time electricity market.  In particular, we are interested in obtaining conditional probability distribution of future LMP, power flow, dispatch levels, and congestion patterns from sample paths of random processes of stochastic parameters such as load and generation processes.  These  sample paths can be generated via Monte Carlo simulation based on stochastic models or by sampling historical traces.

Our approach is one of online learning that acquires sequentially a set of solutions that most frequently appear in Monte Carlo simulations, which allows us to avoid explicit computations of DCOPF solutions.  In particular, we borrow the notion of dictionary learning to explain the ideas behind  the proposed online learning approach to forecasting.  Widely used in the signal processing community, dictionary learning refers to acquiring a dictionary of signal bases to represent a rich class of signals using words (atoms) in the dictionary \cite{kreutz2003dictionary,rubinstein2010dictionaries}.

There are two components of the online learning approach. One is the learning of the underlying stochastic model of the parameter process, the other  the learning of the collection of critical regions that characterizes the solution structure of parametric DCOPF.  Since there is an extensive literature on the former, we  focus here on the problem of learning the structure of parametric DCOPF.


Analogues to dictionary learning in signal processing, the learning process here is also acquiring a dictionary whose words (or atoms)  are critical regions. In particular, each word is associated with  the affine mapping that maps the parameter to the solution of MPLP/MPQP. Therefore, if we treat a realization of the parameter process as a sentence, the dictionary allows us to translate a sentence in the language of  system parameters  to one in the language of LP/QP solutions.

The online dictionary learning process therefore includes (i) checking if a new parameter $\theta$ has already been learned in the past. If not,  (ii) construct a new entry in the dictionary by computing the critical region that contains $\theta$. The problem of checking if there is a critical region in the dictionary that contains $\theta$ can be implemented efficiently by the use of Huffman tree search.   For (ii), the construction of the dictionary is given by  Theorem 1.  The detailed algorithm is summarized in Algorithm 1.

\begin{algorithm}[t]\begin{algorithmic}[1]\caption{Online Dictionary Learning}\label{alg:dcrg}
\STATE{\textbf{Input:} the mean trajectory $\{\bar{d}_1,\bar{d}_2,\cdots, \bar{d}_T\}$ of load and associated (forecast) distributions $\{\mathcal{F}_1, \mathcal{F}_2, \cdots, \mathcal{F}_T\}$.}
\STATE{\textbf{Initialization:} compute the initial critical region dictionary $\mathcal{C}_0$ from the mean load trajectory.}
\FOR{$m=1,\cdots,M$}
\FOR{$t=1,\cdots,T$}\STATE{Generate sample $d_t^m$ and let $\theta_t^m\triangleq (d_t^m, g_{t-1}^m)$.}\STATE{Search $\mathcal{C}_{t-1}^m$ for critical region $C(\theta^m_t)$. }
\IF{$C(\theta^m_t)\in\mathcal{C}_{t-1}^m$}\STATE{Compute $g_t^m$ from the affine mapping $g_{C(\theta^m_t)}^*(\theta)$.}
\ELSE \STATE{Solve $g_t^m$ from DC-OPF (\ref{eqn:dcopf}) using $\theta_m^t$, compute $C(\theta^m_t)$, and update $\mathcal{C}_{t}^m=\mathcal{C}_{t-1}^m\cup\{C(\theta^m_t)\}$.}\ENDIF
\ENDFOR\ENDFOR
\STATE{\textbf{Output:} the critical region dictionary $\mathcal{C}_T^M$.}
\end{algorithmic}\end{algorithm}

\section{Numerical Simulations}
We present in this section two sets of simulation results. The first compares the computation cost of the proposed method with that of direct Monte Carlo simulations.   To this end, we used the 3210 ``Polish network'' \cite{zimmerman2011matpower}.  The second set of simulations focus on probabilistic forecasting.  With this example, we aim to demonstrate the capability of the proposed method in providing joint and marginal distributions of LMPs and power flows, a useful feature not available in existing forecasting methods.

\subsection{General setup}\label{subsec:IV1}
\begin{figure}[H]
\centering
  \includegraphics[width=.38\textwidth]{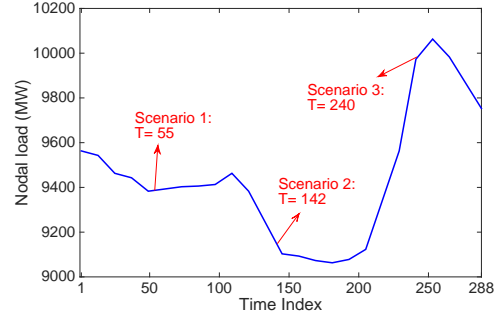}
  \caption{\scriptsize The ``duck curve'' of net load over the different time of the day.}\label{fig:duck}
\end{figure}

We selected the ``duck curve'' \cite{duck} as the expected net load profile as shown in Figure \ref{fig:duck}.  We were particularly interested in three scenarios: Scenario 1 represented a time ($T=55$) when the net load was held steady at the mid range. Scenario 2 ($T=142$) was when the net load was on a downward ramp due to the increase of solar power.  Scenario 3 ($T=240$) was at a time when the net load was at a sharp upward ramp.  The three scenarios represented different operating conditions and different levels of randomness.

The net load---the conventional load offset by renewables---was distributed throughout network.  A renewable generation connected to a bus, say a wind farm,  was modeled as a Gaussian random variable $\Nc(\mu, (\eta \mu)^2)$ with mean $\mu$ and standard deviation $\eta \mu$.  Similar models were used for conventional load forecasts.

Given a forecasting or simulation horizon $T$, the real-time economic dispatch model was a sequence of optimizations with one DCOPF  in each 5 minute interval.   In this model,  the benchmark technique solved a sequence of single period DCOPF models with ramp constraints that coupled the DCOPF solution at time $t$ with that at time $t-1$.   Computationally, the simulation was carried out in a Matlab environment with yalmip toolbox and IBM CPLEX on a laptop with an Intel Core i7-4790 at 3.6 GHz and 32 GB  memory.

\subsection{The 3120-bus System}\label{subsec:3120}
The 3120-bus system (Polish network) defined by MATPOWER \cite{zimmerman2011matpower} was used to compare the computation cost of the proposed method with direct Monte Carlo simulation \cite{Min08}.   The network had 3120 buses, 3693 branches, 505 thermal units, 2277 loads and 30 wind farms. Twenty of the wind farms were located at PV buses and the rest at PQ buses.  For the 505 thermal units, each  unit had upper and lower generation limits as well as a ramp constraint. Ten transmission lines 1, 2, 5, 6, 7, 8, 9, 21, 36, 37 had capacity limits of  275 MW.

The net load profile used in this simulation was the duck curve over a 24 hour simulation horizon.  The total load was at the level of 27,000 MW during morning peak load hours with 10\% of renewables distributed across 30 wind farms.  One large wind farm had rated capacity of  200 MW, 20 midsize wind farms at the rated capacity of 150 MW, and  9 small wind farms at 50-80 MW.  Wind farm $i$ produced Gaussian distributed renewable power $\Nc(\mu_i, (0.03 \mu_i)^2)$.

\begin{figure}[t]
\centering
  \includegraphics[width=.23\textwidth]{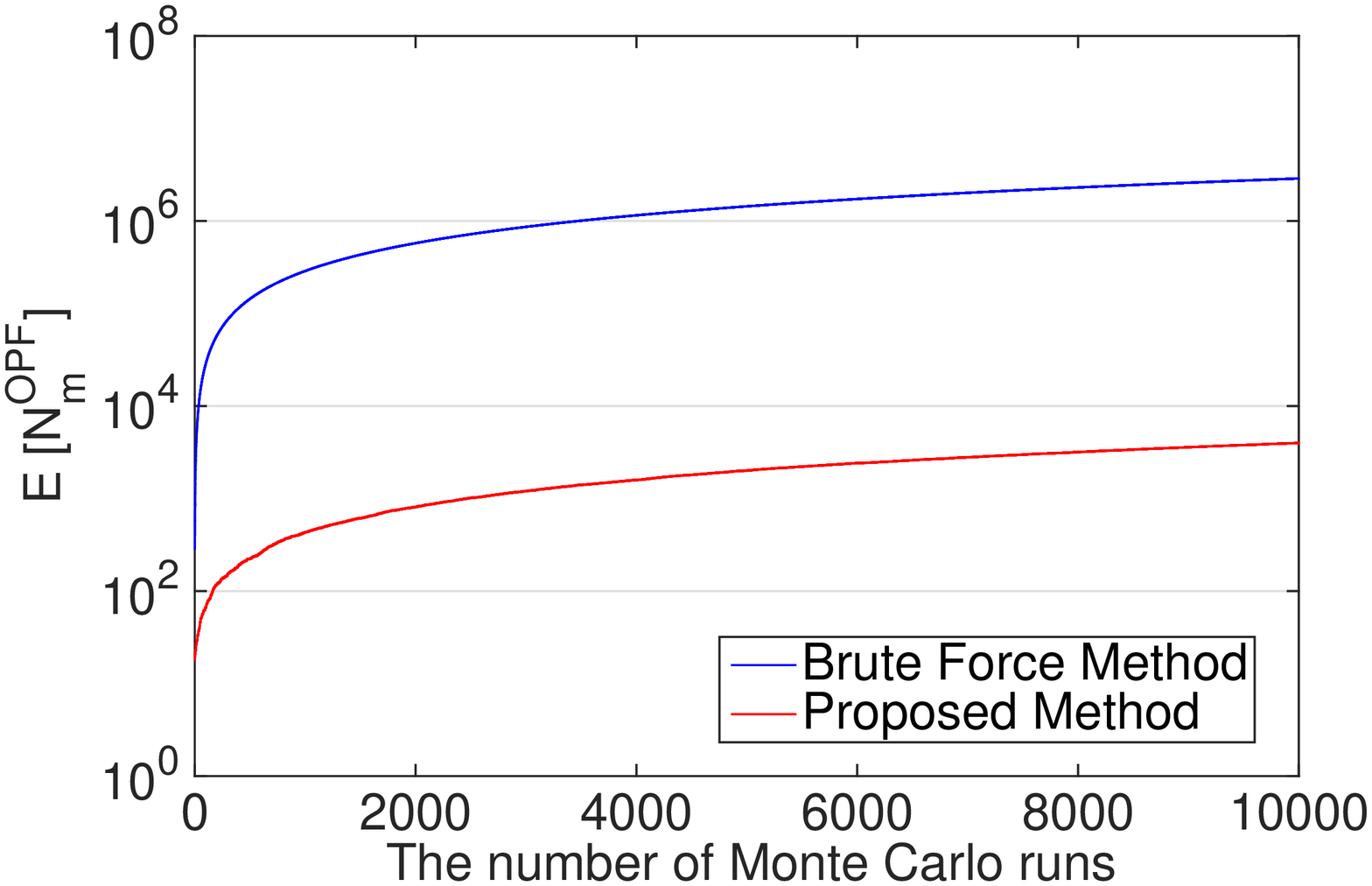}  \includegraphics[width=.23\textwidth]{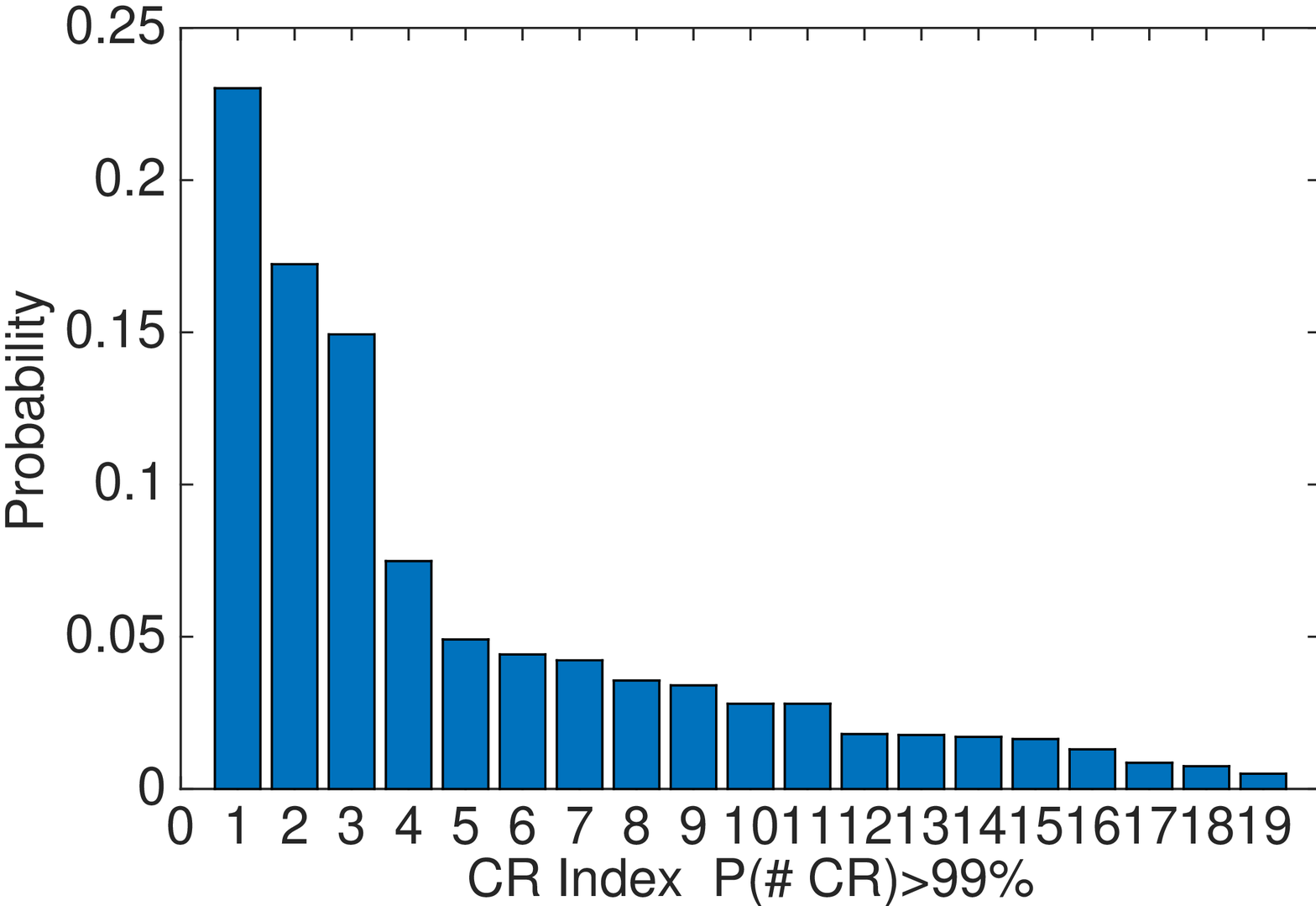}
  \caption{\scriptsize  Left: The expected number of OPF computations vs. the total number of Monte Carlo simulations.  Right: The distribution of the critical regions observed for the proposed method. }\label{fig:OPF}
\end{figure}

The left panel of Figure \ref{fig:OPF} shows the comparison of the computation cost between the proposed approach and the benchmark technique \cite{Min08}.  The two methods obtained identical forecasts, but ODL had roughly three orders of magnitude reduction in the number of DCOPF computations required in the simulation.  This saving came from the fact that only 3989 critical regions appeared in about 2.88 million random parameter samples.   In fact, as shown in the right panel of Figure \ref{fig:OPF},  19 out of 3989 critical regions represented 99\% of all the observed critical regions.

\subsection{The IEEE 118-bus System}\label{subsec:118}
\begin{figure}[H]
\centering
  \includegraphics[width=.40\textwidth]{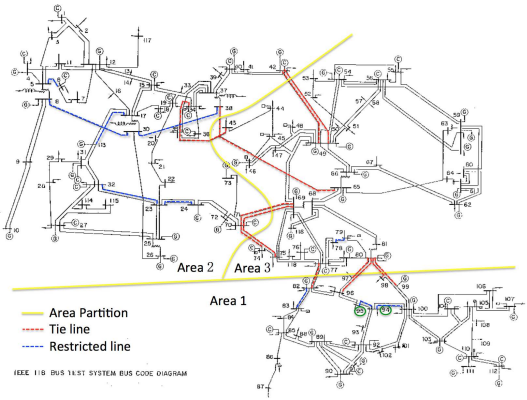}
  \caption{\scriptsize The diagram of IEEE 118-bus system. Blue lines are capacity limited.  Red lines are tie lines.}\label{fig:118}
\end{figure}

The performance of the proposed algorithm was tested on the IEEE 118-bus system \cite{zimmerman2011matpower} shown in Figure \ref{fig:118}. Here the system was partitioned into three subareas.  There were 10 capacity constrained transmission lines (labeled blue) at the maximum capacity of 175 MW. The system included
54 thermal units with ramp limits, 186 branches, and 91 load buses. All of which were connected with  Gaussian distributed load with standard deviation at the level of $\eta=0.15\%$ of its mean. The mean trajectory of the net load again followed the ``duck curve.'' Three scenarios were tested, each included   1000 Monte Carlo runs to generate required statistics.

\subsubsection{Scenario 1: T=55}
\begin{figure}[t]
\centering
  \includegraphics[width=.23\textwidth]{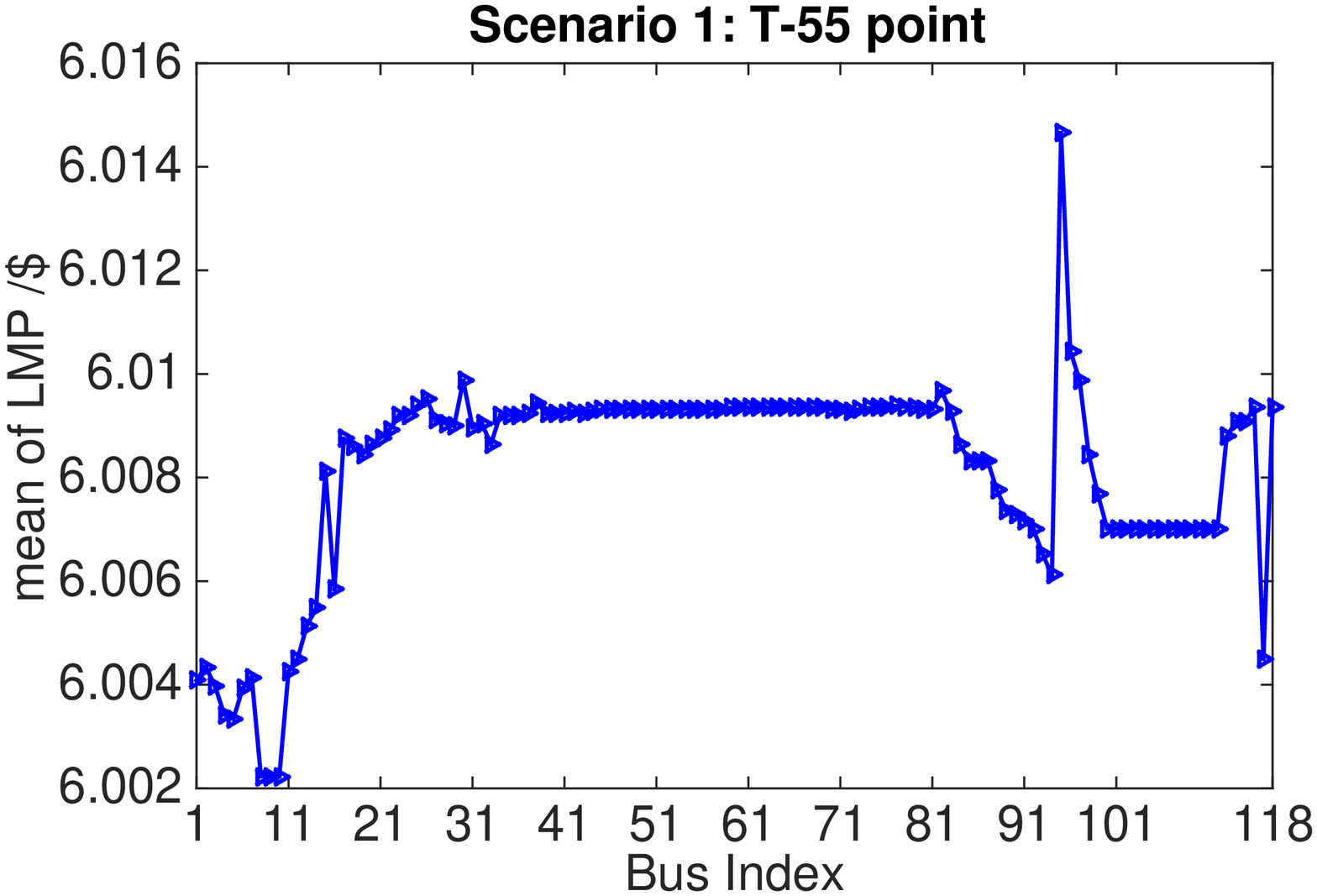} \includegraphics[width=.23\textwidth]{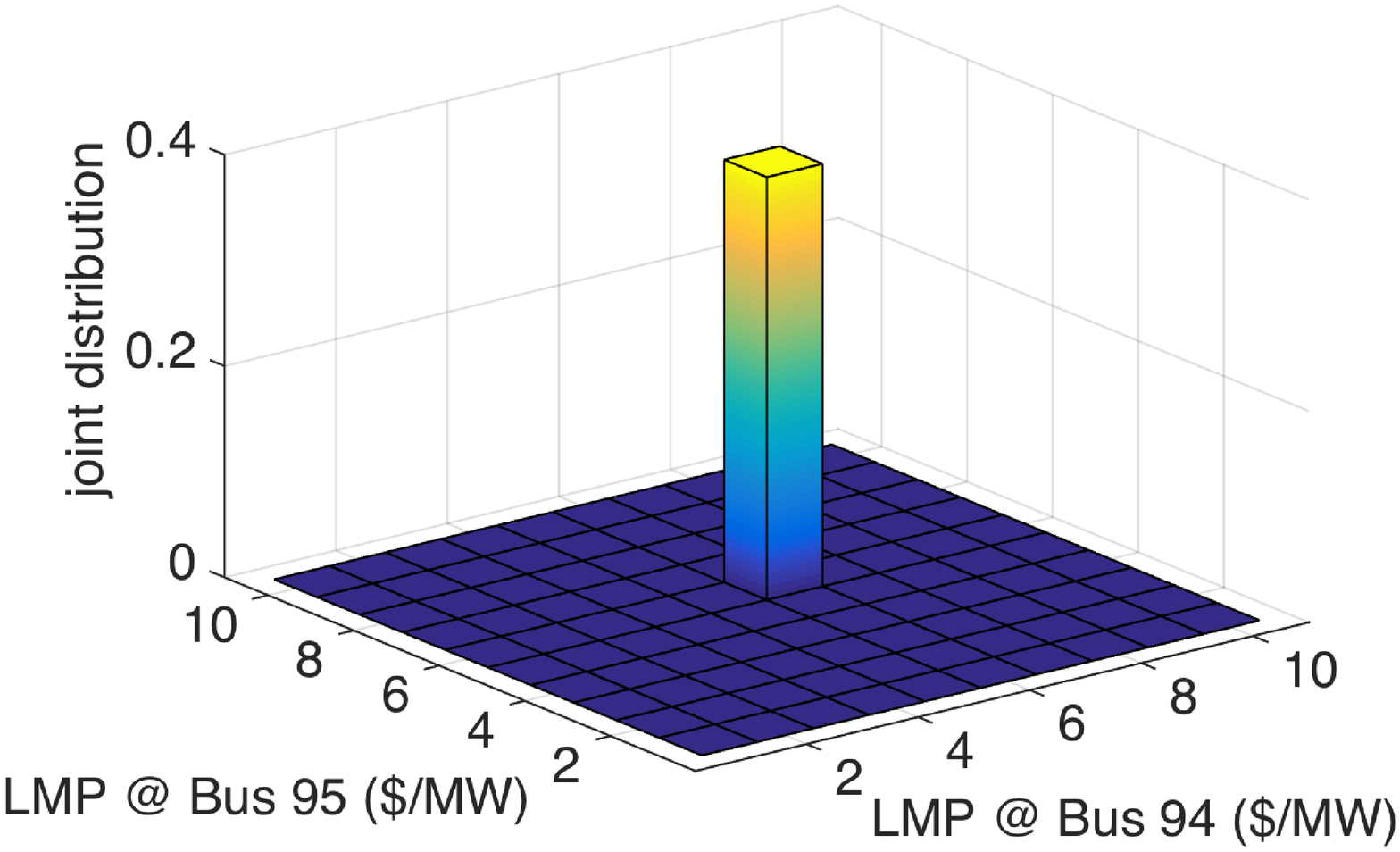}
    \includegraphics[width=.23\textwidth]{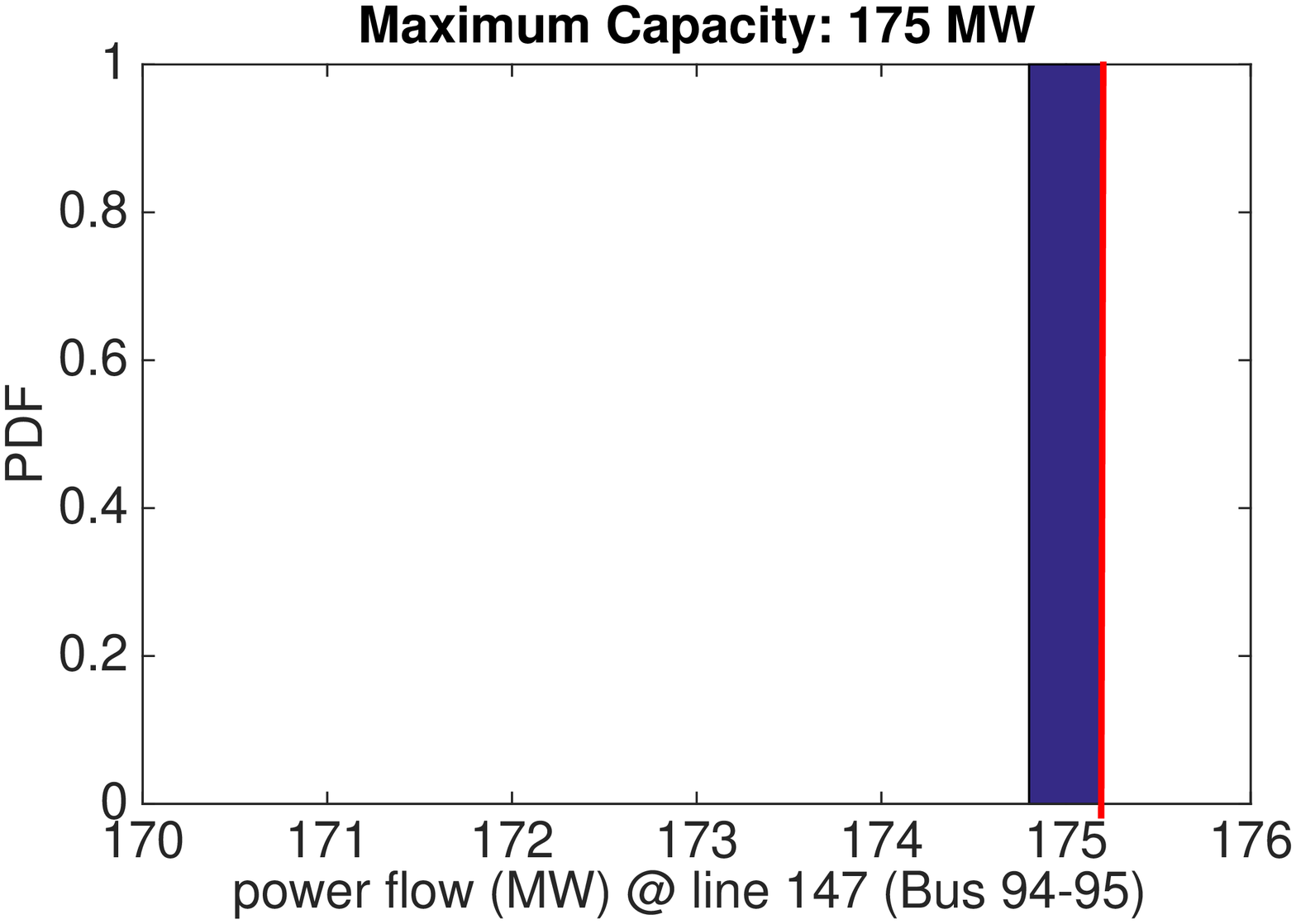} \includegraphics[width=.23\textwidth]{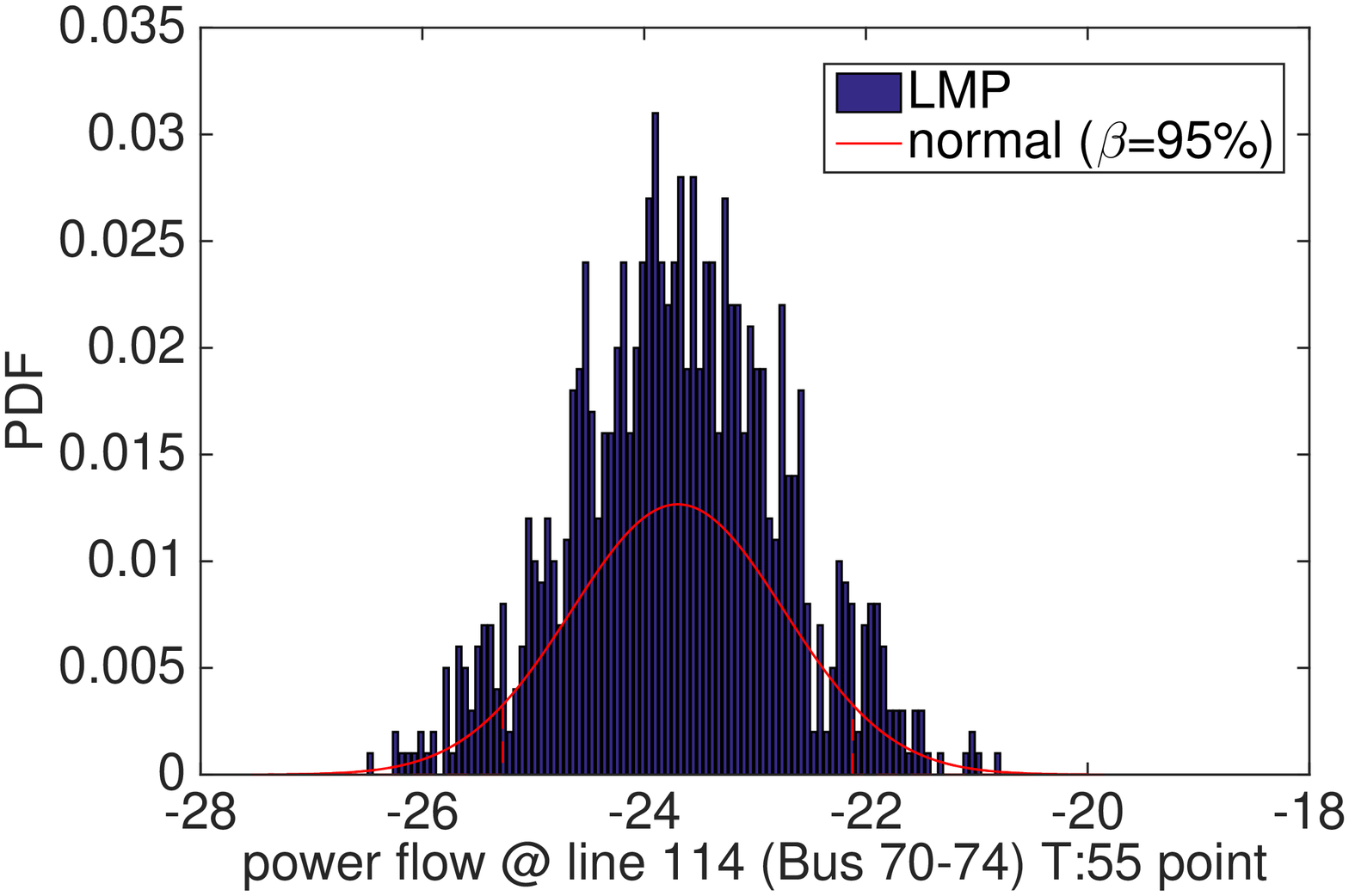}
  \caption{\scriptsize  Top left: The expected LMPs at all buses. Top right: joint LMP distribution at buses 94-95.  Bottom left: power flow distribution on line 147.  Bottom right: power flow distribution on line 114. }\label{fig:T55}
\end{figure}

The first scenario was $T=55$ on the duck curve.  This was a case when the system operated in a steady load regime where the load did not have significant change.  Figure \ref{fig:T55} showed some of the distributions obtained by the proposed technique.  The top left panel showed the average LMP at all buses where the average LMPs were relatively flat with the largest LMP difference appeared  between bus 94 and bus 95.  The top right panel showed the joint LMP distribution at bus 95 and 94.  It was apparent that  the joint distribution of LMP at these two buses was concentrated at a single point mass, which corresponded to the case that all realizations of the random demand fell in the same critical region.  The bottom left panel showed the power flow distribution at line 147 connecting bus 94-95.  As expected, line 147 was congested.   The bottom right panel showed the power flow distribution of line 114, which was one of the tie lines connecting areas 2 and 3. The distribution of power flow exhibited a single mode Gaussian like shape.

\begin{figure}[h]
\centering
  \includegraphics[width=.23\textwidth]{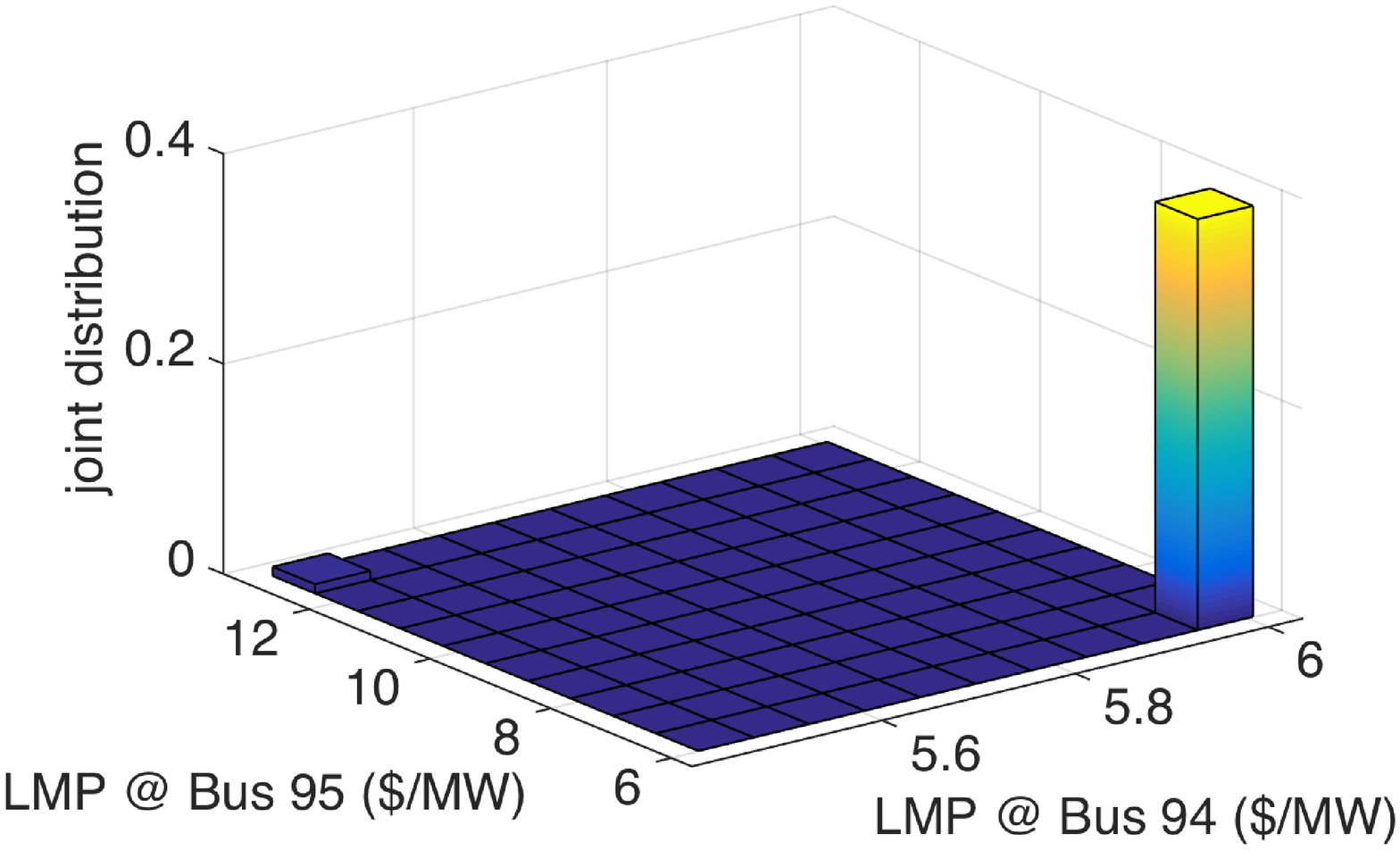} \includegraphics[width=.23\textwidth]{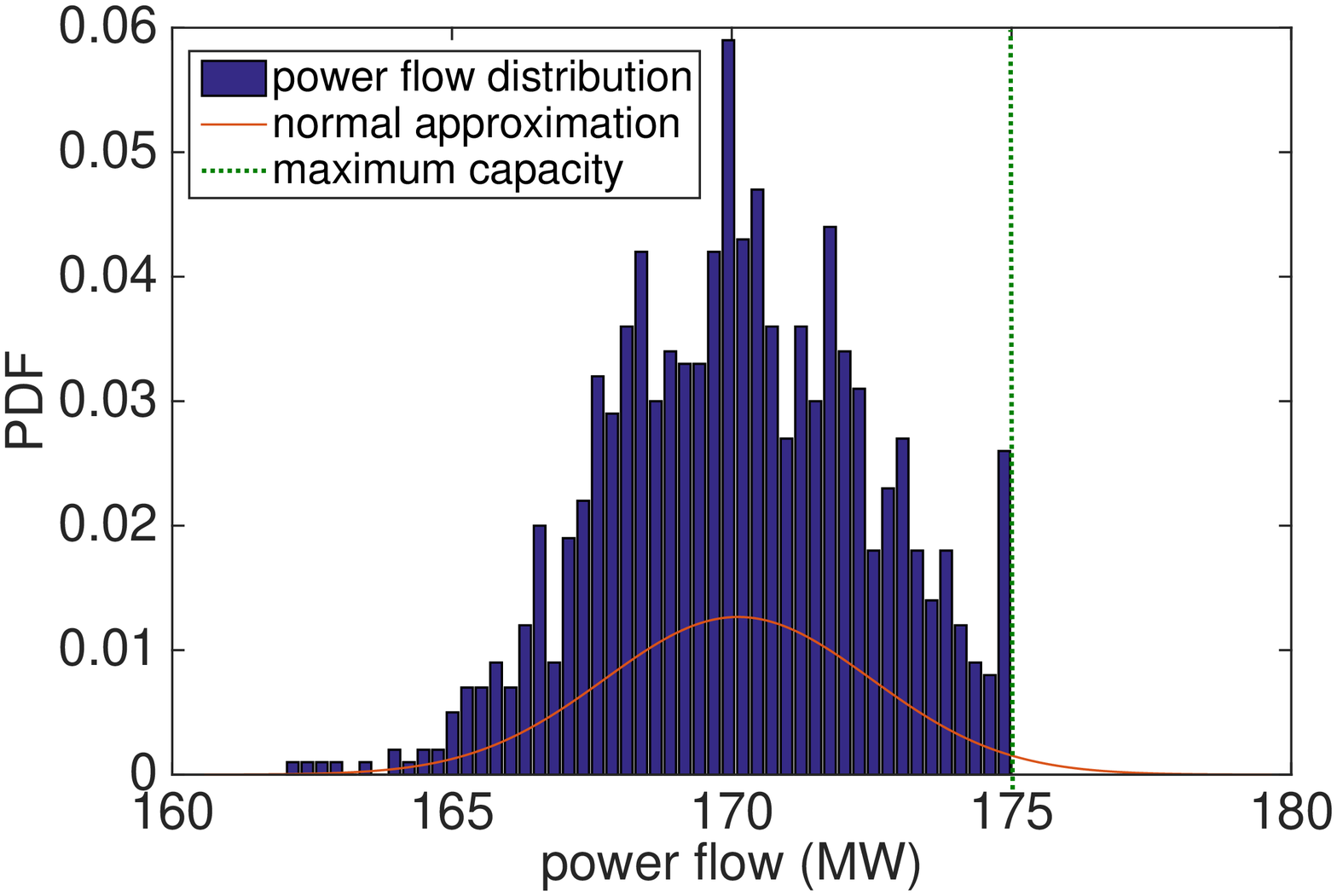}
    \includegraphics[width=.23\textwidth]{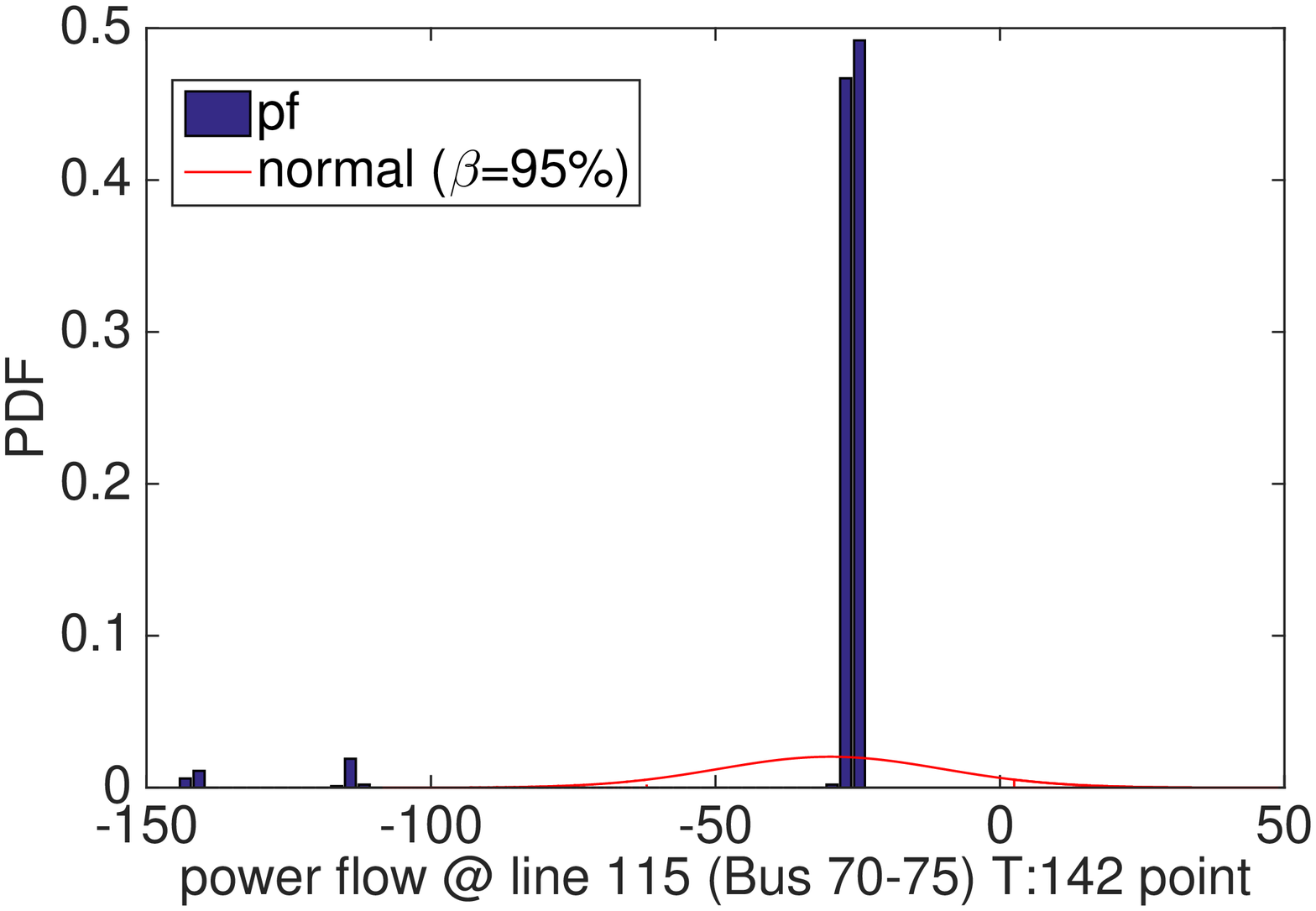} \includegraphics[width=.23\textwidth]{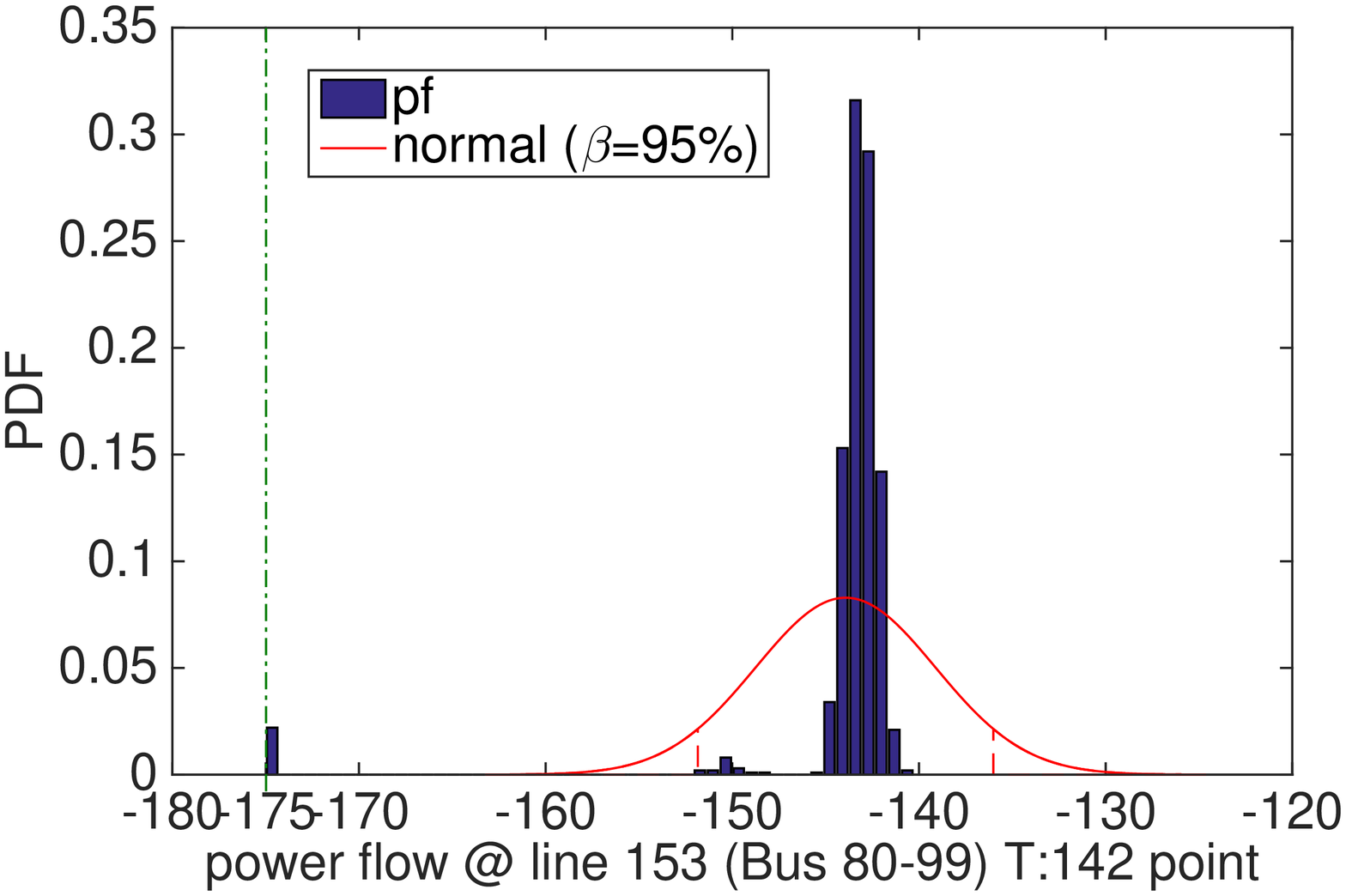}
  \caption{\scriptsize  Top left: joint LMP distribution at buses 94-95.  Top right: power flow distribution on line 147.  Bottom left: power flow distribution on line 115.  Bottom right: power flow distribution on line 153. }\label{fig:T142}
\end{figure}

\subsubsection{Scenario 2: T=142}
The second scenario at T=142 involved a downward ramp.  This was a case when the load crossed boundaries of multiple critical regions.  In Figure \ref{fig:T142}, the top left panel showed the joint probability distribution of LMP at buses 94-95, indicating that the LMPs at these two buses had two possible realizations, one showing small LMP difference with a high probability, the other a bigger price difference with a low probability.
The top right panel showed the power flow distribution on the line connecting bus 94-95.  It was apparent that  the line was congested with non-zero but relatively small probability, which gave rise to the larger price difference between these two buses.  The bottom panels showed the power flow distributions on tie lines 115 and 153.  In both cases, the power flow distribution had three modes, showing little resemblance of Gaussian distributions.

\subsubsection{Scenario 3: T=240}  
\begin{figure}[h]
\centering
  \includegraphics[width=.23\textwidth]{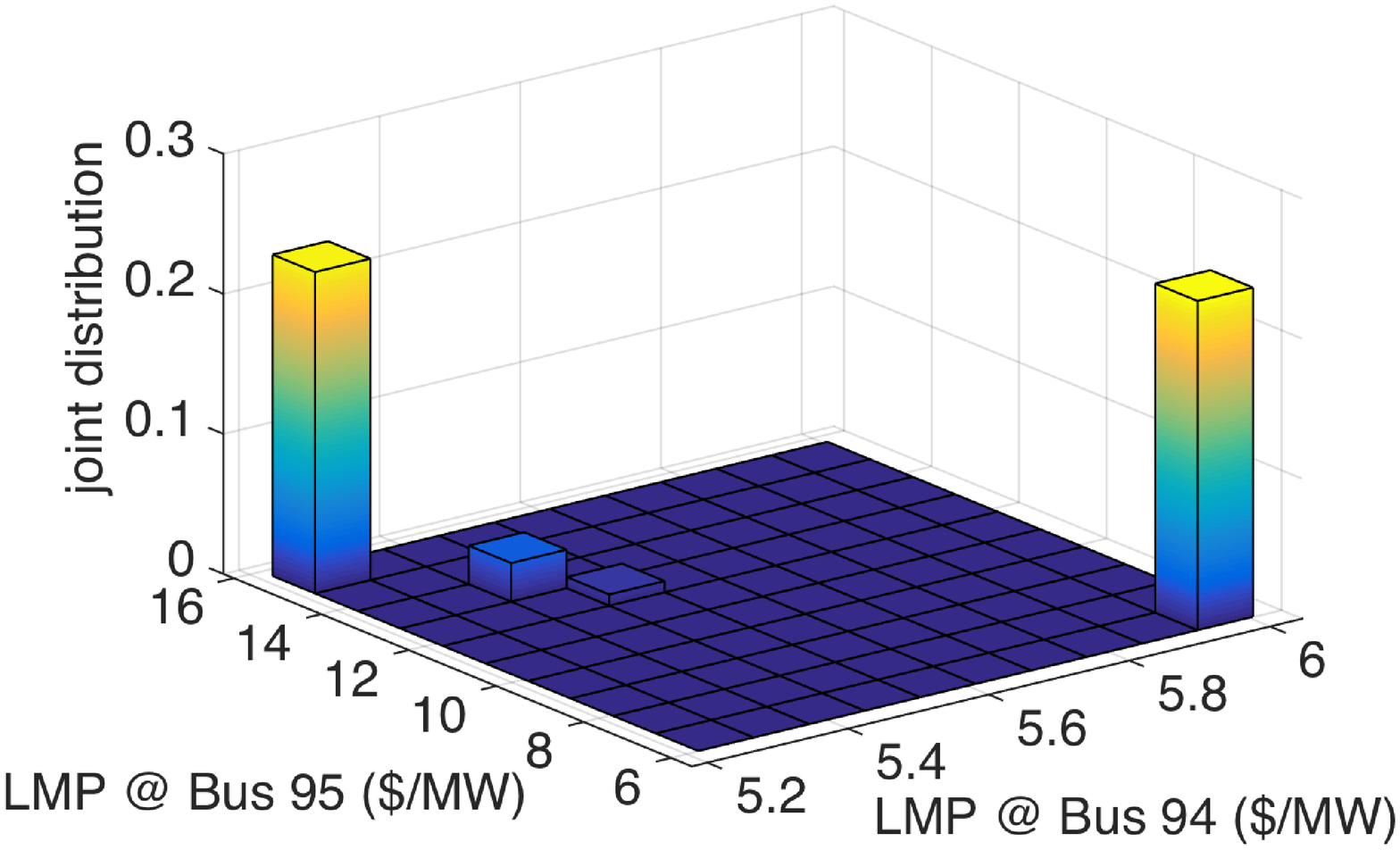} \includegraphics[width=.23\textwidth]{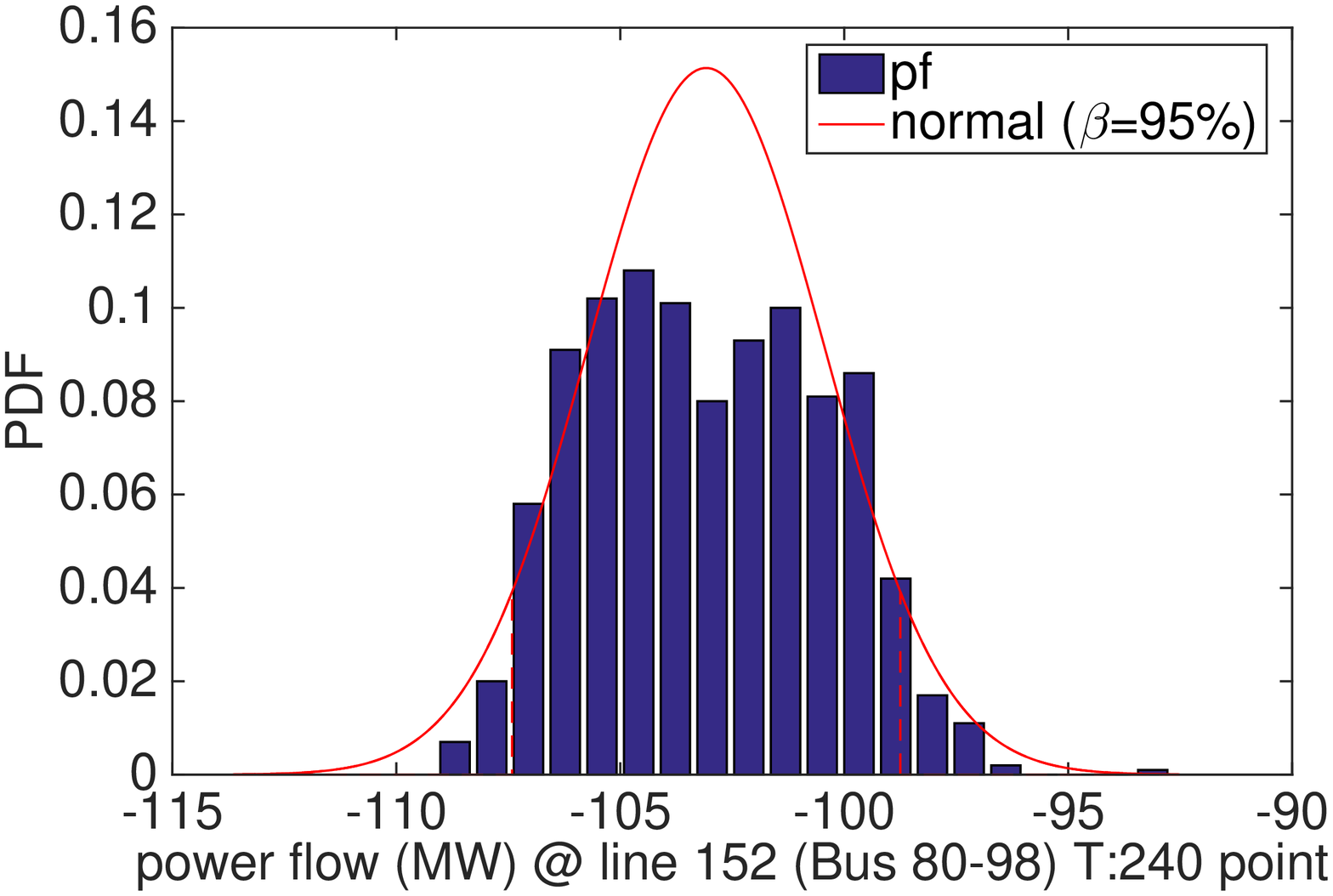}
    \includegraphics[width=.23\textwidth]{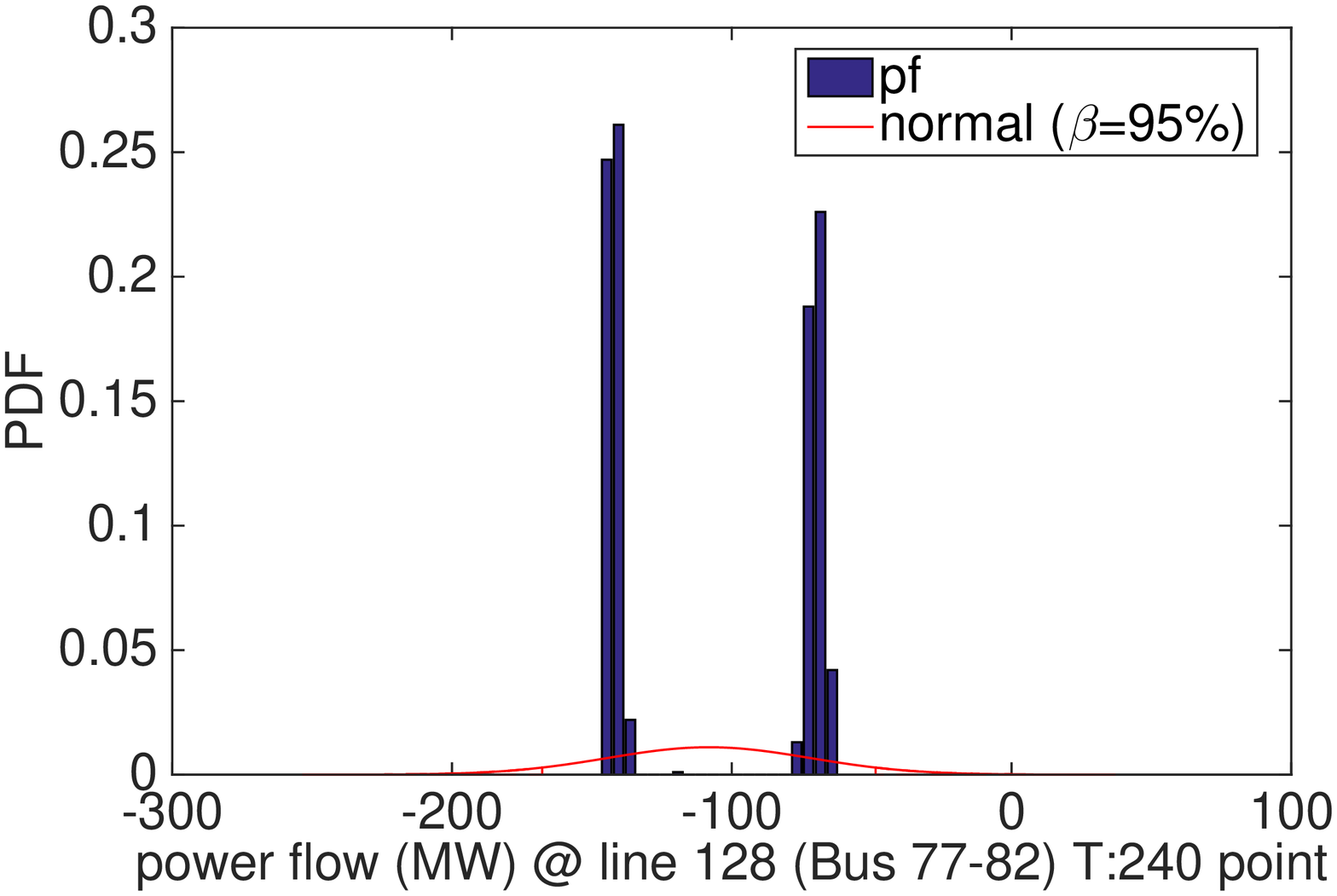} \includegraphics[width=.23\textwidth]{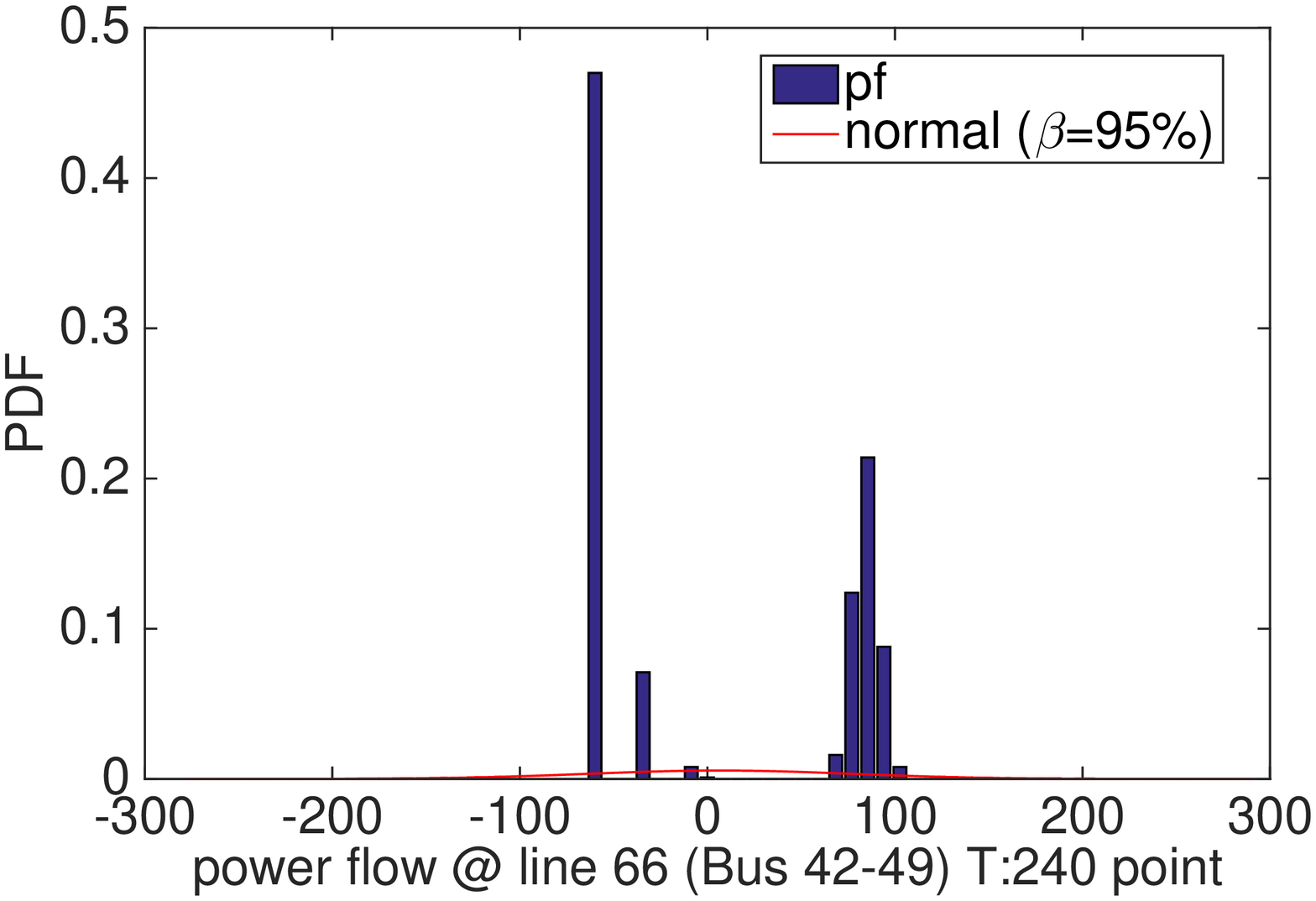}
 \caption{\scriptsize  Top left: joint LMP distribution at buses 94-95.  Top right: power flow distribution on line 152.  Bottom left: power flow distribution on line 128.  Bottom right: power flow distribution on line 152. }\label{fig:T240}
\end{figure}

The third scenario at T=240 involved a steep up ramp at high load levels.  This was also a case when the random load crossed boundaries of multiple critical regions.  In Figure \ref{fig:T240}, the top left panel indicated 4 possible LMP realizations at buses 94-95.  With probability near half that the LMPs across buses 94-95 had significant difference, and the other half the LMPs on these two buses were roughly the same.  The power flow on tie line 152 had a Gaussian-like distribution shown in the top right panel whereas tie line 128 had a power flow distribution spread in four different levels shown in the bottom left panel.  It is  especially worthy of pointing out, from the bottom right panel, that the power flow on line 66 had opposite  directions.

\section{Conclusion}
\label{sec7}
We present in this paper a new methodology of online probabilistic forecasting and simulation of electricity market.  The main innovation is the use of online dictionary learning to obtain sequentially the solution structure of  parametric DCOPF. The resulting benefits are the significant reduction of computation costs and the ability of adapting to changing operating conditions.   Numerical simulations show that, although the total number of critical regions associated with the  parametric DCOPF is very large, only a very small fraction of critical regions appear in a large number of Monte Carlo runs.  This insights highlight the potential of further reducing both computation costs and storage requirements.

\Urlmuskip=0mu plus 1mu\relax
\bibliographystyle{IEEEtran}
{\bibliography{reference4}}

\section*{Appendix: Proof of Theorem~\ref{thm:cr}}
To prove the MPLP case, if $\theta$ is in the same critical region as $\theta_0$, then  $x^*(\theta)$ and  $x^*(\theta_0)$ have the same active/inactive constraints.  This means that
\begin{eqnarray}
\tilde{A}x^*(\theta)-\tilde{b}-\tilde{E}\theta&=&\mathbf{0},\label{pfc:active}\\
\bar{A}x^*(\theta)-\bar{b}-\bar{E}\theta&<&\mathbf{0}.\label{pfc:inactive}
\end{eqnarray}
Because MPLP is neither primal nor dual degenerate,  $\tilde{A}$ has full rank, and
\[x^*(\theta)=\tilde{A}^{-1}(\tilde{b}+\tilde{E}\theta).\]
Substituting $x^*(\theta)$ into (\ref{pfc:inactive}), we have $\theta \in \Cmsc_0$.

Conversely, suppose that $\theta \in \Cmsc_0$.    It can be checked that
\[
x^* \defeq \tilde{A}^{-1}(\tilde{b}+\tilde{E}\theta),~~y^*=y^*(\theta_0)
 \]
 satisfy the KKT condition for being the solution of the MPLP  associated wtih  $\theta$.  Because $x^*$ has the same active/inactive constraints as $x^*(\theta_0)$,  $\theta \in \Cmsc_0$.

For the MPQP case,  suppose that $\theta$ and $\theta_0$ are in the same critical region. Then $x^*(\theta)$ and $x^*(\theta_0)$ have the same active/inactive constraints.  By the KKT condition,
we have
\begin{eqnarray}
H x^*(\theta)+A^\intercal y^*(\theta)&=&\mathbf{0},\label{kkt:stationarity}\\
 \diag(y^{*}(\theta))(Ax^*-b-E\theta) &=&\mathbf{0},\label{kkt:slackness}\\
y^*(\theta)&\geq&\mathbf{0}, \label{kkt:dfeasibility}\\
\tilde{A}x^*(\theta)-\tilde{b}-\tilde{E}\theta&=&\mathbf{0},\label{kkt:active}\\
\bar{A}x^*(\theta)-\bar{b}-\bar{E}\theta&<&\mathbf{0},\label{kkt:inactive}
\end{eqnarray}
where $y^*(\theta)$ is the dual variable and $\diag(y^*(\theta))$ is the diagonal matrix with diagonal entries made of entries of $y^*(\theta)$.
From (\ref{kkt:stationarity}),
\begin{equation}\label{eqn:x(y)}
x^*(\theta)=-H^{-1}A^\intercal y^*(\theta).
\end{equation}
Substituting the result into (\ref{kkt:slackness}), we have
\begin{equation}\label{condition:slack}
\diag(y^{*}(\theta)) (-AH^{-1}A^\intercal y^*-b-E\theta)=\mathbf{0}.
\end{equation}
Let $\bar{y}^*(\theta)$ and $\tilde{y}^*(\theta)$ denote the Lagrange multipliers corresponding to inactive and active constraints respectively. By (\ref{condition:slack}), for inactive constraints, $\bar{y}^*(\theta)=\mathbf{0}$, and for active constraints,
\begin{equation}
\tilde{A}H^{-1}\tilde{A}^\intercal \tilde{y}^*(\theta)+\tilde{b}+\tilde{E}\theta=\mathbf{0}.
\end{equation}
By the non-degeneracy assumption, the rows of $\tilde{A}$ are linearly independent. This implies that $\tilde{A}H^{-1}\tilde{A}^\intercal$ is a square full rank matrix. Therefore
\begin{equation}\label{eqn:y(theta)}
\tilde{y}^*(\theta)=-(\tilde{A}H^{-1}\tilde{A}^\intercal)^{-1}(\tilde{b}+\tilde{E}\theta).
\end{equation}
From (\ref{kkt:dfeasibility}), we have 
\begin{equation}\label{eqn:crd(theta)}
-(\tilde{A}H^{-1}\tilde{A}^\intercal)^{-1}(\tilde{b}+\tilde{E}\theta)\ge  \mathbf{0},
\end{equation}
thus $\theta \in \Pmsc_d$. 
Substituting $\tilde{y}^*(\theta)$ from (\ref{eqn:y(theta)}) into (\ref{eqn:x(y)}), we have
\begin{equation}\label{eqn:x(theta)}
x^*(\theta)=H^{-1}\tilde{A}^\intercal(\tilde{A}H^{-1}\tilde{A}^\intercal)^{-1}(\tilde{b}+\tilde{E}\theta).
\end{equation}
Substituting $x^*(\theta)$ from (\ref{eqn:x(theta)}) in the primal feasibility conditions (\ref{kkt:inactive}),
\begin{equation}\label{eqn:crp(theta)}
\bar{A}H^{-1}\tilde{A}^\intercal(\tilde{A}H^{-1}\tilde{A}^\intercal)^{-1}(\tilde{b}+\tilde{E}\theta)<\bar{b}+\bar{E}\theta,
\end{equation}
thus $\theta\in \Pmsc_p$.  We therefore have $\theta \in \Cmsc_0$.

Conversely, consider $\theta \in \Cmsc_0$.  It can be verified that
\bea
x^* &\defeq&
H^{-1}\tilde{A}^\intercal(\tilde{A}H^{-1}\tilde{A}^\intercal)^{-1}(\tilde{b}+\tilde{E}\theta)\nn\\
\tilde{y}^* &\defeq& -(\tilde{A}H^{-1}\tilde{A}^\intercal)^{-1}(\tilde{b}+\tilde{E}\theta)\nn\\
\bar{y}^* &\defeq& 0\nn
\eea
satisfy the KKT condition, which means that $x^*$ is the solution of (\ref{eqn:mp}).

\end{document}